\documentstyle[12pt]{article}
\newlength{\extraspace}
\newlength{\extraspaces}
\newcommand{\be}{\begin{equation}
\addtolength{\abovedisplayskip}{\extraspaces}
\addtolength{\belowdisplayskip}{\extraspaces}
\addtolength{\abovedisplayshortskip}{\extraspace}
\addtolength{\belowdisplayshortskip}{\extraspace}}
\newcommand{\ee}{\end{equation}}

\newcommand{\ba}{\begin{eqnarray}
\addtolength{\abovedisplayskip}{\extraspaces}
\addtolength{\belowdisplayskip}{\extraspaces}
\addtolength{\abovedisplayshortskip}{\extraspace}
\addtolength{\belowdisplayshortskip}{\extraspace}}
\newcommand{\ea}{\end{eqnarray}}

\begin{document}

\begin{center}
{{\bf Reissner-Nordstr$\ddot{o}$m  anti-de Sitter black holes and
 energy }}
\end{center}
\centerline{ Gamal G.L. Nashed}

\bigskip

\centerline{{\it Mathematics Department, Faculty of Science, Ain
Shams University, Cairo, Egypt }}

\bigskip
 \centerline{ e-mail:nasshed@asunet.shams.eun.eg}

\hspace{2cm}
\\
\\

We show that the tetrad field whose metric gives the
Reissner-Nordstr$\ddot{o}$m  anti-de Sitter black holes gives the
correct value of energy in M\o ller tetrad theory of gravitation.

Mustafa et al. has derived a Reissner-Nordstr$\ddot{o}$m  anti-de
Sitter black holes solution in M\o ller tetrad theory of
gravitation \cite{MO}. Then, a calculations of energy of these
black holes in the spherical polar coordinate have been carried
out. The result of these calculations gives the energy in the form
\be E= M-\displaystyle{Q^2 \over r}+\displaystyle{r^3 \over l^2}.
\ee They discussed this results according to the results of the
calculated energy given in the framework of general relativity. We
have some comments on this discussion. First of all the
calculations of energy that has been done in the spherical polar
coordinate is not right because $P^\mu$ used in these calculations
is not transform as a 4-vector under a linear coordinate
transformation \cite{Mo66}. The calculations will be more accurate
in the Cartesian coordinate \cite{Na1,Vs,Vs1} and we have done
such calculations and obtained the necessary components of the
superpotential of Eq. (22) in Ref.\cite{MO}  \be {{\cal U}_0}^{0
\alpha} \cong {2 n^\alpha \over \kappa
r^3}\left[M-\displaystyle{Q^2 \over 2r}-\displaystyle{r^3 \over
2l^2} \right],\ee Using (2) in (10) of Ref. \cite{MO} we obtain
the energy in the form \be E=m-\displaystyle{q^2 \over
2r}-\displaystyle{r^3 \over 2l^2},\ee which is the correct result
of energy associated with Reissner-Nordstr$\ddot{o}$m  anti-de
Sitter black holes.

\end{document}